\documentclass[conference, letterpaper]{IEEEtran}

\usepackage[left=0.625in,
            right=0.625in,
            top=0.73in,
            bottom=1.1in]{geometry}

\usepackage[cmex10]{amsmath} 
\usepackage{amssymb} 
\usepackage{amsfonts}

\usepackage{subcaption}
\usepackage{graphicx, subcaption}

\usepackage{stfloats}
\usepackage{tikz} 

\def\RR{\mathbb{R}}

\def\CB{\mathcal{C}}

\def\EE{\mathbb{E}}
\def\PP{\mathbb{P}}

\def\eeff{\epsilon_{\text{block}}}

\def\RRR{{\mathcal{R}}}
\def\NNN{{\mathcal{N}}}
\def\FFF{{F_s}}

\def\bb{{b}}
\def\RR{{\rho}}

\DeclareMathOperator*{\argmax}{arg\,max}

\DeclareMathOperator*{\Bin}{Bin}

\newtheorem{theorem}{Theorem}[section]

\newtheorem{lemma}{Lemma}[theorem]
\newtheorem{definition}{Definition}

\begin{document}

\title{Learning to Transmit Over Unknown Erasure Channels with Empirical Erasure Rate Feedback} 

\author{%
\IEEEauthorblockN{Haricharan Balasundaram, Krishna Jagannathan}
\IEEEauthorblockA{Department of Electrical Engineering, IIT Madras\\
                   Chennai, India\\
                   Email: haricharanb@smail.iitm.ac.in, krishnaj@ee.iitm.ac.in}
} 

\maketitle

\begin{abstract}

We address the problem of reliable data transmission within a finite time horizon $T$ over a binary erasure channel with unknown probability of erasure. We consider a feedback model wherein the transmitter can query the receiver infrequently and obtain the empirical erasure rate experienced by the latter. We aim to minimize a \emph{regret} quantity, i.e. how much worse a strategy performs compared to an oracle who knows the probability of erasure, while operating at the same average block error rate. A learning vs. exploitation dilemma manifests in this scenario---specifically, we need to balance between (i) learning the erasure probability with reasonable accuracy and (ii) utilizing the channel to transmit as many information bits as possible. We propose two strategies: (i) a two-phase approach using rate estimation followed by transmission that achieves an $O({T}^{\frac 23})$ regret using only one query, and (ii) a windowing strategy using geometrically-increasing window sizes that achieves an $O({\sqrt{T}})$ regret using $O(\log(T))$ queries.

\end{abstract}

\section{Introduction}

\label{sec:introduction}

Erasure channels are one of the most widely-studied noise models in information theory. We consider a regime where the transmission blocklengths are large but finite, building upon the seminal work of Polyanskiy, Poor, and Verd\'u \cite{finite_block_length_encoding} which establishes achievability and converse bounds for transmission over Discrete Memoryless Channels (DMCs). The problem of finite blocklength encoding has since been extensively studied, with extensions to multi-user \cite{multiuser_it}, multiple-access \cite{multiple_access,multiple_access_2}, fading \cite{fading_channels_1, fading_channels_2, fading_channel_3, impact_finite_blocklength}, and Gaussian channels \cite{gaussian_channels}.

We address the problem of maximizing the information transmitted through a binary erasure channel over a finite time horizon $T$. In a key departure from the usual setting, we assume that the erasure probability is unknown to the transmitter. We assume that there is only \textit{empirical erasure rate} feedback, i.e., whenever the transmitter queries the receiver, the latter conveys only the fraction of bits that have been erased until that point (and not the location of the erased bits). Further, since such queries can be expensive, we are interested in the regime where the number of such queries, $q,$ is `small' in comparison to $T$.

If the receiver provides instantaneous feedback to the transmitter after each symbol reception as studied in \cite{dmcs_with_feedback}, the transmitter can simply retransmit each erased bit. On average, $T(1 - \delta)$ many bits would be received successfully \cite{bec_transmission_elgamal}. However, in our framework this would require $T - 1$ queries to implement, which can be unrealistic and expensive.

In our setting, the key challenge lies in balancing the trade-off between learning the erasure probability with reasonable accuracy and transmitting useful information---thus typifying an exploration-exploitation dilemma. To quantify the performance of a transmission strategy, we define the \emph{regret} of a strategy as the difference between the amount of information transmission achieved by that strategy and the maximum information transmission achieved by an oracle who knows the erasure probability and operates at the same block error rate. Our problem is particularly relevant in the context of IoT and massive Machine Type Communications (mMTC) in cellular networks, where we expect low-resource nodes to transmit a relatively small quantity of information and we expect empirical erasure rate queries to be expensive.

We present two transmission strategies and characterize their regret performance over a long but finite time horizon $T$, under a reasonable step-function approximation of the block-error rate validated through simulations. The first of these is the Estimate-then-Transmit strategy, a two-phase approach wherein we first send a number of pilot bits to estimate the erasure probability, then query for the empirical erasure rate, and then transmit useful information in the remaining time, which achieves an $O(T^\frac 23)$ regret using only one query. The second strategy uses windows with sizes increasing in a geometric progression, which requires $O(\log(T))$ queries and achieves an $O(\sqrt{T})$ regret. Using intuition obtained from the Chernoff bound, we also conjecture the lower bound on the regret if only one query is allowed to be $\Omega(\sqrt{T})$, i.e., no oracle-free policy can obtain a better regret for this problem using only one query.

\noindent \textbf{Related Work.} Our setting is reminiscent of regret minimization in the \textit{multi-armed bandits} problem. Specifically, the Explore-then-Commit strategy studied in \cite{etc_medical, bandits_book} achieves an $O(T^\frac 23)$ regret similar to our Estimate-then-Transmit strategy. Likewise, the $\Omega(\sqrt{T})$ regret lower bound for the instance-independent problem studied in \cite{nonstochastic_mab_lower_bound} aligns with our conjecture on the regret lower bound for the single query case in our problem. However, there appears to be no direct correspondence between the two problems, as the multi-armed bandit problem involves arm-pulls that are independent and identically distributed, whereas in our problem encoding over a block requires handling multiple bits jointly.

\noindent \textbf{Organization of the Paper.} In Section~\ref{sec:preliminaries_and_problem_formulation} we discuss preliminaries and define the problem formally. In Section~\ref{sec:transmission_strategies} we provide theorems regarding the effectiveness of the proposed strategies. Section~\ref{sec:conclusions} provides some further insights and concluding remarks. We relegate all proofs to the Appendices.

\section{Preliminaries and Problem Formulation}

\label{sec:preliminaries_and_problem_formulation}

\subsection{Notation}

We study the Binary Erasure Channel (BEC), in which bits are erased in an i.i.d. manner with erasure probability $\delta$. In our model, $\delta$ is fixed but unknown. Consider a codebook $\CB$ consisting of $|\CB|$ codewords. Each codeword in the codebook is a binary string of length $n$ ($\CB \subseteq {\{ 0, 1\}^n}$). The rate is defined as $\frac kn$, where $k = \log_2(|\CB|)$. The transmitter does not receive feedback on whether an erasure occurred after each transmission, but can query for the empirical erasure rate (the fraction of bits erased) at any point in the transmission.

Let $X \in \CB$ be the input codeword to the BEC, and let $Y \in {\{0, 1, ? \}^n}$ be the corresponding output. For a codeword $X$, there exists a conditional probability mass function $\PP (Y | X)$ that describes the output of the BEC conditioned on the input. Upon observing an output $Y$, we use the Maximum Likelihood Decoder $\hat X = \argmax_{X \in \CB} (\PP(Y | X))$ to predict the input codeword.

The error in our encoder-decoder pair for some $X \in \CB$ is $1 - \PP(\hat X = x | X = x)$. There are two notions of error defined: 

\begin{definition}[Average probability of error]
    \begin{equation}
        \epsilon_{av} = {\frac {1}{|\CB|} \sum_{x \in \CB} \left(1 - \PP(\hat X = x | X = x) \right)}.
    \end{equation}
\end{definition}

\begin{definition}[Maximum probability of error]
    \begin{equation}
        \epsilon_{max} = {\max_{x \in \CB} (1 - \PP(\hat X = x | X = x)))}.
    \end{equation}
\end{definition}

Clearly, $\epsilon_{max} \ge \epsilon_{av}$ and we will consider $\epsilon \equiv \epsilon_{max}$ hereon. It is noted here that $\epsilon$ is a block error (in contrast to $\delta$, which is a bit error). $\epsilon$ depends on the erasure probability $\delta$, the blocklength $n$, and the rate $r$ and thus we denote it as $\epsilon (\delta, n, r)$.

\begin{definition}[${(n, |\CB|, \epsilon)}$-code]
A code with codebook size $|\CB|$, blocklength $n$, and maximum probability of error equal to $\epsilon$. Such a code may not always exist.
\end{definition}

\subsection{Preliminaries on Finite Blocklength BEC}

We recall equivalent versions of the relevant results from the celebrated work of Polyanskiy, Poor, and Verd\'u on encoding with finite blocklengths \cite{finite_block_length_encoding}. There are no known exact expressions for $\epsilon(\delta, n, r)$ and we instead resort to upper and lower bounds.

\begin{theorem} \cite[Theorem 37]{finite_block_length_encoding} For a BEC with erasure probability $\delta$, there exists an $(n, nr, \epsilon)$-code such that
\begin{equation} \epsilon (\delta, n, r) \le \sum_{t = 0}^n {\binom{n}{t}} \delta^t (1 - \delta)^{n - t} 2 ^{-[n (1 - r) - t]^+}
\label{eqn:upper_bound_error},\end{equation}
where $[a]^+ \equiv \max(a, 0)$.
\label{thm:polyanskiy_ub}
\end{theorem}
    
\begin{theorem}\cite[Theorem 38]{finite_block_length_encoding} For a BEC with erasure probability $\delta$ with an $(n, nr, \epsilon)$-code, the error probability satisfies
\begin{equation}
    \epsilon (\delta, n, r) \ge \sum\limits_{t = \lfloor n (1 - r) \rfloor + 1}^n {\binom{n}{t}} \delta^t (1 - \delta)^{n - t} \left (1 - \frac{2^{n(1 - r)}}{2^t} \right).
    \label{eqn:lower_bound_error}
\end{equation}
\label{thm:polyanskiy_lb}
\end{theorem}

\begin{figure}[!ht]
    \centering
    \includegraphics[width=0.80\linewidth]{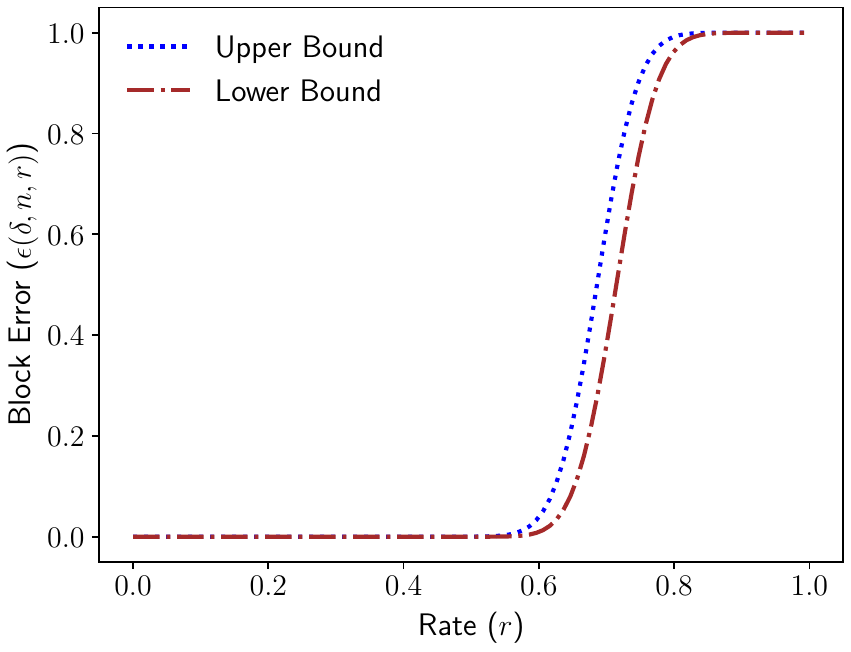}
    \caption{Theorem~\ref{thm:polyanskiy_ub} and~\ref{thm:polyanskiy_lb} for $\epsilon$ for $\delta = 0.3$, $n = 100$.}
    \label{fig:error_lower_upper_bounds}
\end{figure}

As seen in Fig. \ref{fig:error_lower_upper_bounds}, the upper and lower bounds are tight to within a few bits.

\begin{theorem}\cite[Theorem 53]{finite_block_length_encoding} Let $r^* (n, \epsilon)$ be the maximum rate $r$ such that there exists an $(n, n r, \epsilon)$-code. For the BEC with erasure probability $\delta$, we have:
\begin{equation}
    \label{eqn:polyanskiy_opt_rate}
    r^* (n, \epsilon) = (1 - \delta) - \sqrt{\frac{\delta (1 -\delta)}{n}} Q^{-1} (\epsilon) + \frac{O(1)}{n}.
\end{equation}
\end{theorem}

\subsection{Problem Formulation}

In our model, the erasure probability $\delta \in [0, 1]$ is not known \textit{a priori}. Our goal is to find a strategy that maximizes the effective number of bits sent through the channel over a \textit{finite horizon} of $T$ channel uses. The transmitter is allowed only $q$ \textit{queries} of the empirical erasure rate, and this $q$ is typically much smaller compared to $T$.

A \textit{transmission strategy} is defined to be a total of $M$ blocks, with $T_i$ ($1 \le i \le M$) transmission bits in the $i$th block and $\sum_{i = 1}^{M} T_i = T$. We take $\vec T = (T_1, T_2, \dots, T_M)$ and let $C_i = \sum_{j = 1}^{i} T_j$ be the number of transmission till the end of block $i$. 

At the end of the $i$th block, the strategy receives the empirical erasure rate feedback. More formally, if there were $K_i \sim \Bin(C_i, \delta)$ erasures during transmissions up to the end of block $i$, the strategy receives $\hat \delta_i = \frac{K_i}{C_i}$. For the $(i+1)$th block, the strategy has to decide the rate of transmission $\rho_{i + 1}(C_i, \hat \delta_i)$. It is clear that such a strategy requires $q = M - 1$ empirical erasure rate queries.

\begin{figure}[!ht]
    \centering

    \begin{tikzpicture}[scale=0.66]
        \draw (-2,0) rectangle (10,1.6);

        \draw (-0.5,0) -- (-0.5,1.6);   
        \draw (2,0) -- (2,1.6);         
        \draw (4,0) -- (4,1.6);         

        \node at (-1.25,1.15) {$T_1$};
        \node at (0.75,1.15) {$T_2$};
        \node at (3,0.8) {$\dots$};
        \node at (7,1.15) {$T_M$};

        \node at (-1.25,0.45) {$\rho_1$};
        \node at (0.75,0.45) {$\rho_2(T_1,\hat{\delta}_1)$};
        \node at (7,0.45) {$\rho_M(T - T_{M},\hat \delta_{M-1})$};

        \node at (4,-0.3) {$T$ bits};
    \end{tikzpicture}

    \caption{Transmission strategy}
    \label{fig:schematic_windowing}
\end{figure}
A schematic for a general strategy is displayed in Fig.~\ref{fig:schematic_windowing}.

Note that in the $i$th block, the $\rho_i$'s are random variables. For a particular realization of $\rho_i = r_i$, the block transmission may be a success or a failure, dictated by the block error probability $\epsilon(\delta, T_i, r_i)$. Since $\rho_i$ itself is a random variable, the probability of a block error, ${\eeff}_i$ would be 
\begin{IEEEeqnarray}{rCl}
    {\eeff}_i &=& \EE[\epsilon(\delta, T_i, \rho_i(C_{i - 1}, \hat \delta_{i - 1}))] \label{eqn:eeff_expression} \\
    &=& \sum_{k = 0}^{C_{i - 1}} \delta^k (1 - \delta)^{C_{i - 1} - k} \epsilon \left(\delta, T_i, \rho_i \left(C_{i - 1}, \frac k{C_{i - 1}} \right) \right). \nonumber
\end{IEEEeqnarray}.

Next, note that $C_{i} \rho_i$ represents the number of information bits transmitted in block $i$. The expected number of `successfully-decoded' information bits transmitted in the $i$th block would be:
\begin{IEEEeqnarray}{rCl}
    \NNN_i &=&  \EE[T_i \rho_i \cdot \epsilon(\delta, T_i, \rho_i(C_{i - 1}, \hat \delta_{i - 1}))].
\end{IEEEeqnarray}

Any strategy tries to maximize $\NNN = \sum_{i = 1}^M \NNN_i$. The average block error incurred by any strategy is taken to be the weighted average of the individual block errors:
\begin{equation}
    \eeff = \frac{\sum_{i = 1}^M {\eeff}_i T_i}{T}.
    \label{eqn:block_error}
\end{equation}
This choice of $\eeff$ is motivated in the next subsection.

The performance of a strategy with a certain expected number of successfully-decoded information bits, $\NNN(\eeff)$, is compared against an oracle who is transmitting at the same $\eeff$ as our strategy. Such an oracle would transmit at a rate $r^*(T, \eeff)$ given by \eqref{eqn:polyanskiy_opt_rate} and thus the expected number of successfully-decoded information bits sent by this oracle is equal to:
\begin{equation}
    \NNN^o(\eeff) \equiv T \cdot r^*(T, \eeff) \cdot (1 - \eeff).
    \label{eqn:optimal_oracle}
\end{equation}

This leads us to the notion of the \textit{regret} (denoted by $\RRR (\eeff)$), which quantifies the performance gap between the oracle and a given strategy. For a particular strategy with an average probability of block error $\eeff$, $\RRR(\eeff) = \NNN^o(\eeff) - \NNN(\eeff)$. Our aim is to formulate strategies to maximize $\NNN(\eeff)$, or equivalently, to minimize $\RRR (\eeff)$.

\subsection{Choice of $\eeff$}

In this section, we motivate the choice to weight the ${\eeff}_i$'s in~\eqref{eqn:block_error} by their respective blocklengths $T_i$. Consider a strategy with $M$ blocks where each block is transmitted at rate $\rho_i = 1$. Block
$i$ is decoded successfully only if none of its $T_i$ transmitted bits are erased, i.e.
\begin{equation}
    {\eeff}_i = 1 - (1-\delta)^{T_i}.    
\end{equation}
The expected number of successfully decoded information bits is therefore
\begin{equation}
\NNN = \sum_{i=1}^M T_i \cdot 1 \cdot (1-{\eeff}_i) = \sum_{i=1}^M T_i(1-\delta)^{T_i}.    
\end{equation}
By the definition of $\eeff$, note that we get \begin{equation}
    \NNN = T (1 - \eeff).
\end{equation}
Thus, when all the rates are $1$, $\eeff$ has the interpretation that it is the number of successful information bits transmitted as a fraction of the time horizon.

\section{Transmission Strategies and Analyses}

\label{sec:transmission_strategies}

\subsection{Estimate-then-Transmit}

In this strategy, we divide our horizon into the estimation phase and the transmission phase as depicted in Fig. \ref{fig:schematic_exploration_exploitation}, with $T_1 = T_e$ and $T_2 = T_t$.

In the estimation phase, we send only zeroes in the channel and no meaningful information is transmitted. The rate is taken to be $0$. At the end of this phase, we query the empirical erasure rate, $\hat \delta = \hat \delta_1 = \frac{K}{T_e}$, where the subscript is dropped for convenience. We take $\NNN_1 = 0$ and $\eeff = {\eeff}_2$ since this is an upper bound on the actual average block error. 

\begin{figure}[!ht]
    \centering
    \begin{tikzpicture}[scale=0.8]
        \draw (0,0) rectangle (10,1);       
        \draw (3,0) -- (3,1);        
        \node at (1.5,0.5) {$T_e$ bits};
        \node at (6.5,0.5) {$T_t$ bits};

        \node at (1.5,1.3) {Estimation Phase};
        \node at (6.5,1.3) {Transmission Phase};

        \node at (5,-0.3) {$T$ bits};

    \end{tikzpicture}

    \caption{Estimate-then-Transmit strategy}
    \label{fig:schematic_exploration_exploitation}
\end{figure}

In the transmission phase, we send information at the rate $\rho = \RR_2 = \max(0, 1 - \hat \delta - \bb) = \max(0, 1 - \frac{K}{T_e} - \bb)$. The parameter $\bb \ge 0$ indicates the rate back-off. Intuitively, a non-zero backoff could be useful, because for large $T$, transmitting at a rate slower than $1 - \delta$ is nearly error-free while transmitting at a rate faster than $1 - \delta$ results in a block error with probability close to $1$. In other words, a pessimism in the rate would incur a modest penalty that is linear in the rate backoff $\bb$, but an optimism in the rate (i.e., a negative backoff $\bb$) would result in a catastrophic increase in the block error.

The value of $\eeff$ stems from~\eqref{eqn:eeff_expression}. Since we do not know of a tractable expression for $\epsilon(\delta, T_t, r)$, we can utilize the bounds from \eqref{eqn:upper_bound_error} and \eqref{eqn:lower_bound_error} to get upper and lower bounds on $\eeff (\bb)$. We omit such expressions in the interest of space. The value of $\NNN$ is as below:
\begin{IEEEeqnarray}{L}
\NNN (\eeff, \delta, T, T_e) = \EE \left[T_t \cdot (1 - \epsilon(\delta, T_t, \RR)) \cdot \RR \right] \nonumber \\
= T_t \cdot \sum_{k = 0}^{T_e} \binom{T_e}{k} \delta^{k} (1 - \delta)^{T_e -k} \cdot \left(1 - \epsilon \left(\delta, T_t, r \right) \right) \cdot r, \label{eqn:NNN_eff_exact_expr} \end{IEEEeqnarray}
where $r = \max(0, 1 - \frac{k}{T_e} - \bb)$.
We utilize the bounds from \eqref{eqn:upper_bound_error} and \eqref{eqn:lower_bound_error} for ${\epsilon(\delta, T_t, r)}$ to obtain \eqref{eqn:fraction_lower_bound} and \eqref{eqn:fraction_upper_bound} below.

\begin{figure*}[!b]
\hrulefill
\begin{equation}
    \NNN (\eeff, \delta, T, T_e) \ge T_t \cdot \sum_{k = 0}^{T_e} \binom{T_e}{k} \delta^{k} (1 - \delta)^{T_e -k} \cdot \left(1 - \sum_{t = 0}^{T_t} {\binom{T_t}{t}} \delta^t (1 - \delta)^{T_t - t} 2 ^{-[T_t s_k - t]^+} \right) \cdot \max\left(0, 1 - \frac{k}{T_e} - \bb \right), \label{eqn:fraction_lower_bound}
\end{equation}
\begin{equation}
    \NNN (\eeff, \delta, T, T_e) \le T_t \cdot \sum_{k = 0}^{T_e} \binom{T_e}{k} \delta^{k} (1 - \delta)^{T_e -k} \cdot \left(1 - \sum\limits_{t = \lfloor T_t s_k \rfloor + 1}^{T_t} {\binom{T_t}{t}} \delta^t (1 - \delta)^{T_t - t}  \left (1 - \frac{2^{T_t s_k}}{2^t} \right) \right) \cdot \max \left(0, 1 - \frac{k}{T_e} - \bb \right), \label{eqn:fraction_upper_bound}
\end{equation}

\vspace{-0.2cm}where $s_k = \min \left(1, \frac{k}{T_e} + \bb \right)$.
\end{figure*}

Our aim is to maximize $\NNN(\eeff, \delta, T, T_e)$ as a function of $T_e$ for a given $T$ and $\delta$. However, the expressions for $\eeff$ as well as the bounds on $\NNN(\eeff)$ in  \eqref{eqn:fraction_lower_bound} and \eqref{eqn:fraction_upper_bound} are complicated, and optimizing with respect to $T_e$ appears to be intractable. To glean intuition regarding the performance of our strategy for large but finite $T$ and obtain expressions for the leading-term behavior, we approximate the error function for large $T$ as described below. We later justify the validity of our approximation using numerical computations.

As the value of the blocklength $n$ increases, the plot in Fig.~\ref{fig:error_lower_upper_bounds} begins to resemble a step function; the probability of block error would be approximately $1$ for $r > (1 - \delta)$ and approximately $0$ for $r < (1 - \delta)$. Using this insight, we set the approximate expression for error in terms of rate as below:
\begin{equation}
\epsilon(\delta, n, r) = 
\begin{cases} 
0 & \text{for } 0 \leq r \le (1 - \delta), \\ 
1 & \text{for } r > (1 - \delta)
\end{cases}.
\end{equation}

We refer to this as the \textit{step function approximation}. This expression for the error is independent of $n$ and depends only on $\delta$ and $r$. Under this approximation, we first characterize $\eeff(\bb)$ from \eqref{eqn:eeff_expression}.
\begin{theorem}
    Under the step function approximation, for a backoff $\bb$ and empirical erasure rate feedback obtained for $T_e$ bits, the error incurred in a blocklength $T_t$ is $Q\left(\bb \sqrt{\frac{T_e}{\delta (1 - \delta)}} \right) + O \left(\frac{1}{\sqrt{T_e}} \right)$, where $Q$ denotes the tail distribution of the standard normal distribution.
    \label{thm:epsilon_eff(B)}
\end{theorem}
The proof of this result follows from the Berry-Esseen theorem \cite{berry_esseen}. We note that the $O \left(\frac{1}{\sqrt{T_e}} \right)$ error due to the Berry-Esseen theorem is small and thus set $\eeff \equiv Q \left(\bb \sqrt{ \frac{T_e}{\delta (1 - \delta)}} \right)$ hereon. Thus, choosing the backoff and $T_e$ determines the value of $\eeff$ for a given channel and we consider $\NNN$ and $\RRR$ as functions of $\bb$ instead.
Our strategy is thus compared against an oracle whose $\NNN^o(\eeff)$ is given by \eqref{eqn:optimal_oracle}.

Next, we provide a theorem that characterizes $\NNN$ as a function of parameters $T_e$ and $\eeff$ for a given $T$ and $\delta$.

\begin{theorem} \label{thm:exp_exploration} Under the step function approximation, for the Estimate-then-Transmit strategy, only one empirical erasure rate query is required to achieve
\begin{IEEEeqnarray}{L}
\label{eqn:NNN_expression}
    \NNN(\bb, \delta, T, T_e) = \nonumber \\ T_t \left((1 - \delta - \bb) (1 - \eeff (\bb)) -\sqrt{\frac{\delta (1 - \delta)}{2 \pi T_e}} e^{-\frac{(Q^{-1}(\eeff(\bb)))^2}{2}} \right) \nonumber \\ \hspace{6cm} + O \left(\frac{T_t}{T_e^\frac 32}\right),
\end{IEEEeqnarray}
where $T_t = T - T_e$.
\end{theorem}

Next we provide a theorem that characterizes the optimum $T_e$ for a given $\eeff$.

\begin{theorem} \label{thm:NNN_expression_maximization}
The value of $T_e$ that maximizes \eqref{eqn:NNN_expression} for a given $\eeff$ and $T$, $T_e^*$, is the solution to
\begin{multline}
\hspace{-0.5cm} \frac{(T+T_e^*)\sqrt{\delta(1-\delta)}}{2(T_e^*)^{3/2}}
\left(
Q^{-1}(\eeff)(1-\eeff)
+\frac{e^{-\frac{1}{2}[Q^{-1}(\eeff)]^2}}{\sqrt{2\pi}}
\right)
\\=(1-\delta)(1-\eeff).
\label{eqn:exploration_exploitation_soln}
\end{multline}
Equation \eqref{eqn:exploration_exploitation_soln} has a unique solution $T_e^*$ with $T_e^* = \Theta(T^{\frac 23})$ and thus the regret satisfies $\RRR(\eeff, \delta, T, T_e^*) = O(T^\frac 23)$.
\end{theorem}

As argued already, for different values of $\bb$ and hence $\eeff$, we would expect different values of $\NNN(\bb)$. The value of $\bb$ which maximizes $\NNN(\eeff)$ is given by the following theorem.
\begin{theorem}
The value of $\bb$ that maximizes $\NNN$ for a given $T_e$ occurs at $\eeff^*$ given by
\begin{equation}
        (1 - \delta) \cdot e^{-\frac{(Q^{-1} (\eeff^*))^2}{2}} \sqrt{\frac{T_e}{2 \pi \delta (1 - \delta)}}  = (1 - \eeff^*). \label{eqn:optimum_NNN}
\end{equation}
\label{thm:optimum_NNN}
\end{theorem}
Simultaneously solving \eqref{eqn:exploration_exploitation_soln} and \eqref{eqn:optimum_NNN} enables us to get $\eeff$ and $\bb$ for a given $T$ and $\delta$.

\begin{figure*}[t]
    \centering
    \begin{subfigure}[t]{0.32\linewidth}
        \centering
        \includegraphics[width=\linewidth]{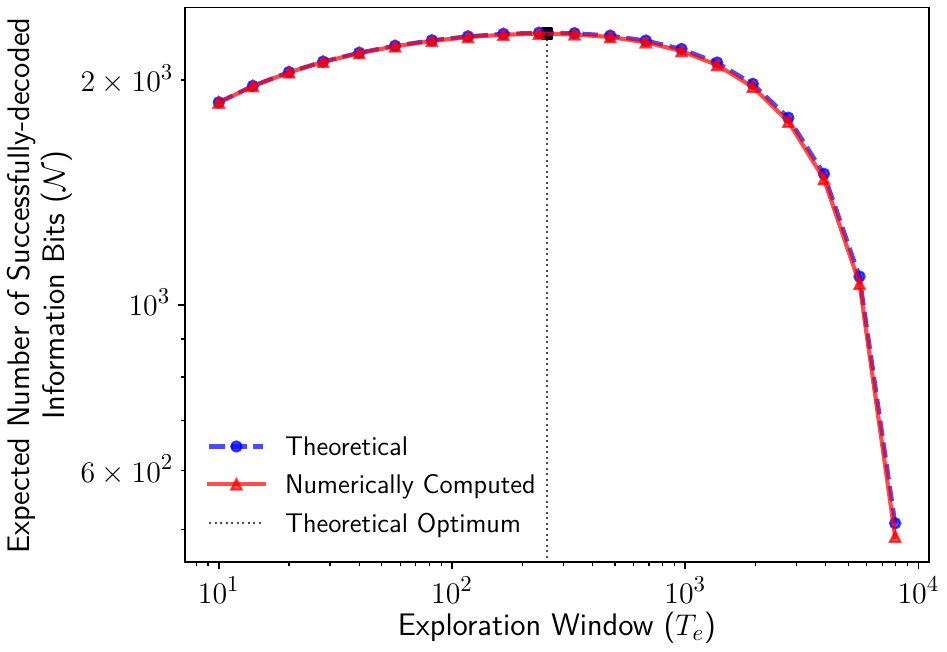}
        \caption{$\NNN$ vs $T_e$}
        \label{fig:optimum_Te}
    \end{subfigure}
    \hfill
    \begin{subfigure}[t]{0.32\linewidth}
        \centering
        \includegraphics[width=\linewidth]{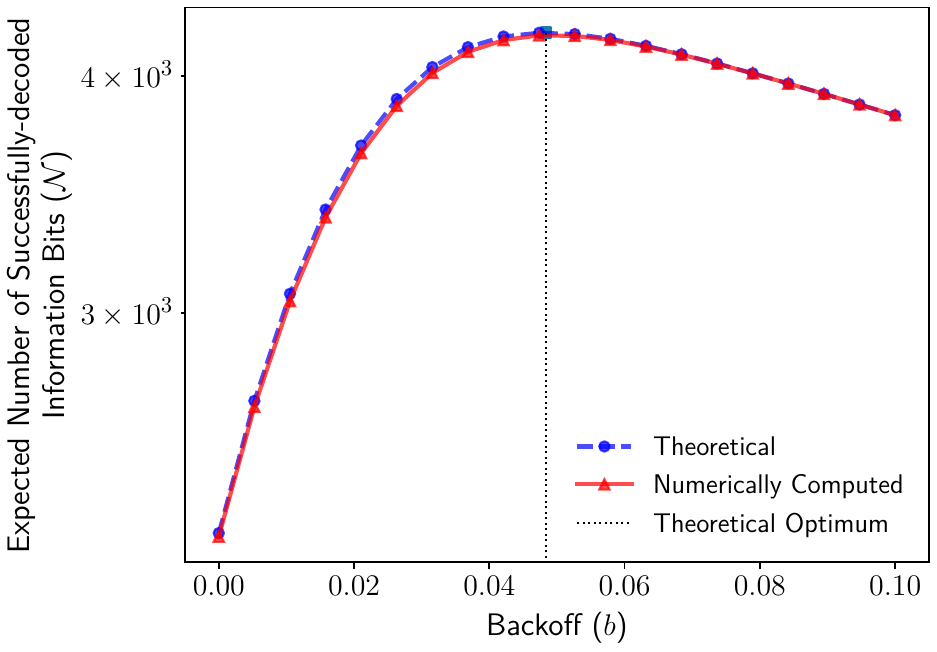}
        \caption{$\NNN$ vs $\bb$}
        \label{fig:N_vs_backoff}
    \end{subfigure}
    \begin{subfigure}[t]{0.29\linewidth}
        \centering
        \includegraphics[width=\linewidth]{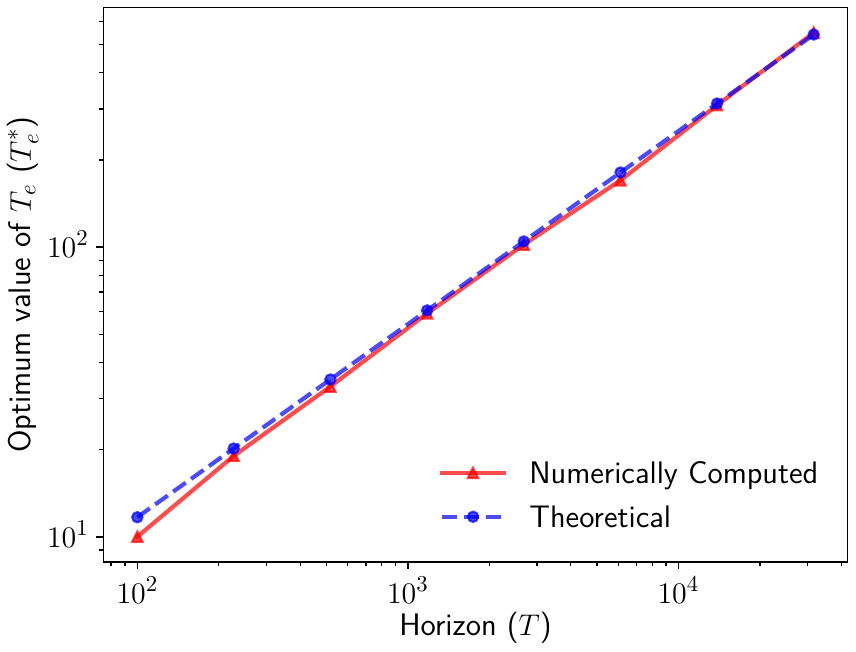}
        \caption{$T_e^*$ vs $T$}
        \label{fig:T_e_opt_vs_T}
    \end{subfigure}

    \caption{All plots have $\delta = \frac 12$ for the estimate-then-transmit strategy. (a) and (b): Expected number of successfully-decoded information bits as a function of $T_e$ and $b$ respectively for $T = 10000$. (c): Optimum estimation phase length vs $T$.}
    \label{fig:combined_figures}
\end{figure*}

\subsubsection*{Simulations} We first verify that the expression given in Theorem~\ref{thm:exp_exploration} is correct. We plot the expected number of successfully-decoded information bits sent, $\NNN$, against $T_e$ in Fig.~\ref{fig:optimum_Te} and against the backoff $\bb$ in Fig.~\ref{fig:N_vs_backoff} for $\delta = \frac 12$, $T = 10000$ and $\eeff = \frac 12$ (i.e., $\bb = 0$). The theoretical optimum values from Theorem~\ref{thm:NNN_expression_maximization} and Theorem~\ref{thm:optimum_NNN} are also plotted in the respective figures. We see that the theoretical and the numerically computed values agree closely in both cases, and that the optimum value also matches.

To further verify Theorem~\ref{thm:NNN_expression_maximization}, we plot the optimum exploration window $T_e^*$ against $T$ for $\delta = 0.5$ and $\eeff = 0.5$ in Fig.~\ref{fig:T_e_opt_vs_T}. We see that the value of $T_e^*$ calculated theoretically agrees with the value computed numerically. The straight line has a slope of approximately $\frac 23$, as expected theoretically.

\subsection{Geometric Windowing}

Motivated by the backoff in the previous section, we consider windowed strategies for a total of $T$ blocks, where we take $\RR_i = \max(0, 1 - \hat \delta_{i - 1} - \bb_i)$ for a suitable backoff $\bb_i$. We take $\eeff = {\eeff}_i \forall 1 \le i \le M$, i.e. the block error probabilities in each of the blocks are the same. The backoff $\bb_i$ would be such that the block error rate is exactly $\eeff$, i.e., $b_i = \sqrt{\frac{\delta (1 - \delta)}{\sum_{j = 1}^{i - 1} T_j}}Q^{-1} (\eeff)$.

The $i$th block ($i > 1$) thus behaves like the transmission phase of an estimate-then-transmit strategy with $T_e = C_{i - 1}$ and $T_t = T_i$. Using Theorem \ref{thm:exp_exploration} for one block and summing that quantity over all the $M$ blocks, it is possible to characterize the value of $\NNN$ for any set of blocklengths $(T_1, T_2, \dots, T_M)$.

The geometric windowing strategy is a specific case of the windowing strategy wherein we consider windows with sizes $T_i = 2^{i - 1}$ for $1 \le i \le M$. We assume $T$ is such that $T = 2^{M} - 1$ for some $M$. In such a case, we would require $O(\log(T))$ empirical erasure rate queries. The $\NNN$ achieved by this strategy is given by Theorem~\ref{thm:geometric_windowing}.

\begin{theorem} Under the step function approximation, for the Geometric Windowing strategy, $O(\log T)$ empirical erasure rate queries are sufficient to achieve $\NNN (\eeff, \delta, M, \vec T)$ characterized by the following upper and lower bounds:
\begin{multline*}
    \NNN \le T (1 - \delta) (1 - \eeff) - \frac{\sqrt{(T + 1) \delta (1 - \delta)}}{\sqrt{2} - 1} f(\eeff),
\end{multline*}
\begin{multline*}
    \NNN \ge T (1 - \delta) (1 - \eeff) - \frac{\sqrt{2(T + 1) \delta (1 - \delta)}}{\sqrt{2} - 1} f(\eeff),
\end{multline*}
\begin{equation*}
    \text{where } f(\eeff) = Q^{-1}({\eeff}) (1 - \eeff) + \frac{e^{-\frac{(Q^{-1} (\eeff))^2}{2}}}{\sqrt{2 \pi}}.
\end{equation*}
\label{thm:geometric_windowing} 
\end{theorem}
Theorem~\ref{thm:geometric_windowing} gives us the $\Theta(\sqrt{T})$ regret bound for this strategy.

\subsection{Numerical Comparison of Strategies}

\begin{figure}[!ht]
    \centering
    \includegraphics[width=0.77\linewidth]{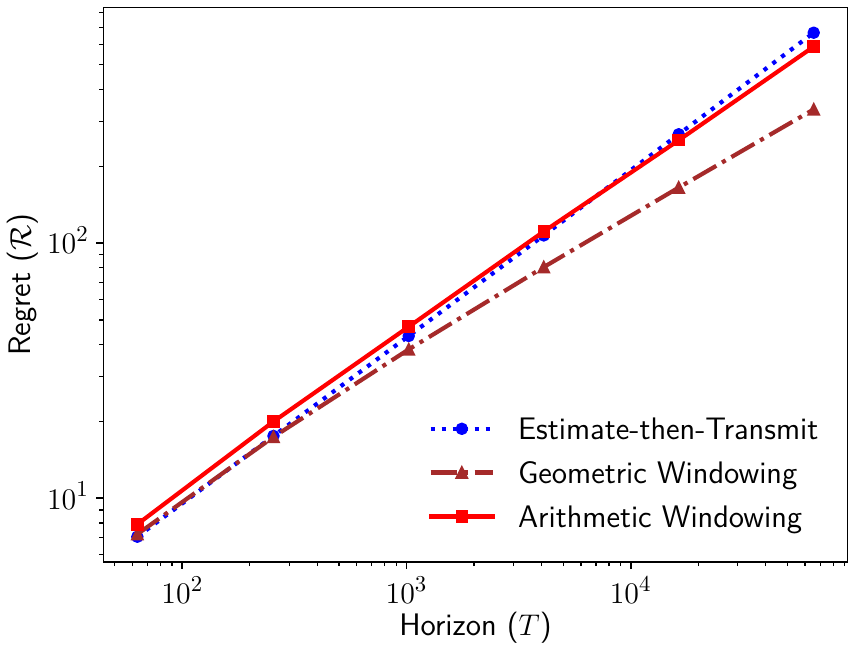}
    \caption{Regret vs. Horizon for the Estimate-then-Transmit and Geometric windowing strategies for $\delta = 0.5$ and $\eeff = \frac 12$}
    \label{fig:comparison}
\end{figure}

We know theoretically that the Geometric Windowing strategy with an $O(\sqrt{T})$ regret performs better than the Estimate-then-Transmit strategy with an $O(T^\frac 23)$ regret for $T_e^*$, but uses a greater number of queries. An `Arithmetic Windowing` strategy with window sizes increasing starting from $1$ in an arithmetic progression such that the number of queries is the same as that in the geometric windowing strategy is expected to have an intermediate performance between the two strategies. The numerically-computed regrets are shown in Fig.~\ref{fig:comparison} for $\delta = \frac 12$ and $\eeff = \frac 12$ (i.e., $\bb = 0$). The Geometric windowing strategy has a slope of roughly $\frac 12$ for sufficiently large values of $T$ while the Estimate-then-Transmit strategy has a slope of roughly $\frac 23$.

\section{Conclusions}
\label{sec:conclusions}

We addressed the problem of reliable data transmission over a binary erasure channel with unknown erasure probability and empirical erasure rate feedback. Using well-known results from finite blocklength information theory, we proposed two strategies: the Estimate-then-Transmit strategy and the Geometric Windowing strategy, both of which utilize the empirical erasure rate feedback to decide the transmission rate.

We conjecture an $\Omega(\sqrt{T})$ regret lower bound for the single query case, reasoning as follows. The Chernoff bound imposes a $\Theta \left (\frac 1 {\sqrt{n}} \right)$ error on the empirical erasure rate after sending $n$ bits. This would necessitate a rate pessimism of $\Theta \left (1/{\sqrt{n}} \right)$ for each $n$, leading to a regret of $\Theta\left ((T - n) / {\sqrt{n}} \right)$ which cannot be made smaller than $\Omega(\sqrt T)$. Formalizing this policy-independent lower bound is an immediate future direction. 

This paper opens up several interesting avenues for future research on the topic of ``learning to transmit'' information theory. Firstly, it seems intuitively clear that as the number of empirical erasure rate queries increases, the regret decreases. However, we do not yet have a fundamental characterization of the `number of queries versus regret' frontier. Secondly, we could consider other noise models, such as the binary symmetric and Gaussian channels. We could also extend our framework to channels with Markovian memory, and potentially utilize techniques from Partially Observed Markov Decision Processes (POMDPs).
Multi-user versions (Broadcast/MAC/Relay etc.) of ``learning to transmit" could also pique future interests.

\newpage

\section*{Acknowledgments}

The authors acknowledge V. Arvind Rameshwar for insightful discussions and valuable feedback.

\bibliographystyle{IEEEtran}
\bibliography{refs}

\clearpage

\appendices

\onecolumn

\section{Proof of Theorem~\ref{thm:epsilon_eff(B)}}
\label{appendix:appendix_1}

\begin{IEEEproof} Let $K \sim \text{Bin}(T_e, \delta)$. For $0 \le K \le T_e(\delta - \bb)$, $1 - \delta \le r \le 1 - \bb$, so we are in error. If $F_K$ is the CDF of $\Bin(T_e, \delta)$,
\begin{IEEEeqnarray}{rCl}
    \eeff(\bb) &= & F_K(T_e(\delta - \bb)) \\
    &\ge & \Phi \left(\sqrt{T_e} \frac{ \delta - \bb - \delta}{\sqrt{\delta (1 - \delta)}} \right) - C \frac{(\delta^2 + (1 - \delta)^2)}{\sqrt{T_e \delta (1 - \delta)}}
    \\ &=& Q \left(\bb \sqrt{ \frac{T_e}{\delta (1 - \delta)}} \right) - C \frac{(\delta^2 + (1 - \delta)^2)}{\sqrt{T_e \delta (1 - \delta)}},
\end{IEEEeqnarray}
using the Berry-Esseen Theorem, where $C$ is a fixed positive constant. \cite{berry_esseen}.

Similarly, $\eeff(\bb) \le Q \left(\bb \sqrt{ \frac{T_e}{\delta (1 - \delta)}} \right) + C \frac{(\delta^2 + (1 - \delta)^2)}{\sqrt{T_e \delta (1 - \delta)}}$. This gives us the desired bound. \end{IEEEproof}

\section{Proof of Theorem~\ref{thm:exp_exploration}}
\label{appendix:appendix_2}

Let $\FFF(\bb, \delta, T, T_e) = \frac{\NNN(\bb, \delta, T, T_e)}{T_t}$, where $T_t = T - T_e$. Under our approximation, we are not in error when $T_e(\delta - \bb) < K \le T_e$, where $K \sim \Bin(T_e, \delta)$ as defined previously.
\begin{IEEEeqnarray}{rCl}
        \FFF(\bb, \delta, T, T_e) &=& \EE \left[(1 - \epsilon(\delta, T_t, \RR)) \cdot \RR \right] \\
        & = & \sum_{k = \lceil T_e (\delta - \bb) \rceil}^{T_e} \max \left(0, 1 - \frac{k}{T_e} - \bb \right) f_K(k)
        \\ &\overset{(a)}{=}& \sum_{k = \lceil T_e (\delta - \bb) \rceil}^{\lfloor T_e (1 - \bb) \rfloor} \left(1 - \frac{k}{T_e} - \bb \right) f_K(k) \label{eqn:F_s_expression},
\end{IEEEeqnarray}
where $(a)$ follows since the summation term is non-zero only if $1 - \frac{k}{T_e} - b \ge 0$.

\begin{lemma}[Lower bound on $F_s$]
\begin{equation}
    \FFF(\bb, \delta, T, T_e) \ge -\sqrt{\frac{\delta (1 - \delta)}{2 \pi T_e}} e^{-\frac{\bb^2 T_e}{2 \delta (1 - \delta)}}  + (1 - \delta - \bb) \cdot (1 - \eeff) + O\left(\frac{1}{T_e^\frac 32}\right)
\end{equation}
\end{lemma}

\begin{IEEEproof} We see that the proof is as below, where $(a)$ comes from bounding the binomial PMF with the normal CDF as in \cite{Auld2024}, $(b)$ comes from bounding the summation with an integral, $(c)$ comes from performing a variable substitution $u = \frac{k - T_e \delta}{\sqrt{T_e \delta (1 - \delta)}}$, and $(d)$ and the further steps come from evaluating and simplifying the integral.

\begin{IEEEeqnarray}{rCl}
        \FFF(\bb, \delta, T, T_e) &\overset{(a)}{\ge}& \sum_{k = \lceil T_e (\delta - \bb) \rceil}^{\lfloor T_e (1 - \bb) \rfloor} \left(1 - \frac{k}{T_e} - \bb \right) \cdot \left(\frac{1}{\sqrt{2 \pi T_e \delta (1 - \delta)}} e^{-\frac{(k - T_e \delta)^2}{2 T_e \delta (1 - \delta)}} - \frac{C}{T_e \delta (1 - \delta)} e^{-\frac{|k - T_e \delta|}{\sqrt{T_e \delta (1 - \delta)}}} \right) \\
        &\overset{(b)}{\ge}& \int_{T_e (\delta - \bb)}^{T_e (1 - \bb)} \left(1 - \frac{k}{T_e} - \bb \right) \left(\frac{1}{\sqrt{2 \pi T_e \delta (1 - \delta)}} e^{-\frac{(k - T_e \delta)^2}{2 T_e \delta (1 - \delta)}} - \frac{C}{T_e \delta (1 - \delta)} e^{-\frac{|k - T_e \delta|}{\sqrt{T_e \delta (1 - \delta)}}} \right) dk + O\left(\frac{1}{T_e^\frac 32}\right) \\
        &\overset{(c)}{\ge}& \int_{-\bb \sqrt{\frac{T_e}{\delta (1 -\delta)}}}^{(1 - \delta - \bb)\sqrt{\frac{T_e}{\delta (1 -\delta)}}} \left(-\sqrt{\frac{\delta (1 - \delta)}{T_e}} u + 1 - \delta - \bb \right) \frac{e^{-\frac{u^2}{2}}}{\sqrt{2 \pi}} du + O\left(\frac{1}{T_e^\frac 32}\right) \\
        &\overset{(d)}{\ge}& \left[\sqrt{\frac{\delta (1 - \delta)}{2 \pi T_e}} e^{-\frac{u^2}{2}} + (1 - \delta - \bb) \cdot \Phi
        \left(u \right) \right]_{-\bb \sqrt{\frac{T_e}{\delta (1 -\delta)}}}^{(1 - \delta - \bb)\sqrt{\frac{T_e}{\delta (1 -\delta)}}} + O\left(\frac{1}{T_e^\frac 32}\right) \\
        &=& \sqrt{\frac{\delta (1 - \delta)}{2 \pi T_e}} (e^{-\frac{(1 - \delta - \bb)^2 T_e}{2 \delta (1 - \delta)}} - e^{-\frac{\bb^2 T_e}{2 \delta (1 - \delta)}}) \nonumber \\ & & + (1 - \delta - \bb) \cdot \left(\Phi \left((1 - \delta - \bb)\sqrt{\frac{T_e}{\delta (1 - \delta)}} \right) - \Phi \left(-\bb \sqrt{\frac{T_e}{\delta (1 - \delta)}} \right) \right) + O\left(\frac{1}{T_e^\frac 32}\right) \\
        &=& \sqrt{\frac{\delta (1 - \delta)}{2 \pi T_e}} ( - e^{-\frac{\bb^2 T_e}{2 \delta (1 - \delta)}})  + (1 - \delta - \bb) \cdot \Phi \left(\bb \sqrt{\frac{T_e}{\delta (1 - \delta)}} \right) + O\left(\frac{1}{T_e^\frac 32}\right),
\end{IEEEeqnarray}

Finally, we substitute the value of $\eeff$ from Theorem~\ref{thm:epsilon_eff(B)} to get the desired result.\end{IEEEproof}

An analogous upper bound can also be obtained, which we omit in the interest of space. Using the lemma, we can obtain a tight bound for $N(\bb, \delta, T, T_e)$, which proves our theorem.

\section{Proof of Theorem~\ref{thm:NNN_expression_maximization}}
\label{appendix:appendix_3}

For maximizing \eqref{eqn:NNN_expression}, we first neglect the $O \left(\frac{T_t}{T_e^{\frac 32}} \right)$ terms. Note that while $\eeff$ is kept constant, $b$ is still not a constant. \eqref{eqn:NNN_expression} can be re-written as
\begin{equation}
    \NNN = (T - T_e) \cdot \left[(1 - \delta) (1 - \eeff) - \sqrt{\frac{\delta(1 - \delta)}{T_e}} (Q^{-1} (\eeff) (1 - \eeff) + \frac{1}{\sqrt{2 \pi}} e^{- \frac{(Q^{-1} (\eeff))^2}{2}})\right].
\end{equation}
Differentiating this with respect to $T_e$, we obtain
\begin{IEEEeqnarray}{rCl}
\frac{d\mathcal N}{dT_e} = -(1-\delta)(1 - \eeff) + \frac{T+T_e}{2T_e^{3/2}} \sqrt{\delta(1-\delta)} \left[ Q^{-1}(\eeff)(1-\eeff) + \frac{1}{\sqrt{2\pi}} e^{\left(- \frac{\left(Q^{-1}(\eeff) \right)^2}{2} \right)} \right]
\end{IEEEeqnarray}
Simplifying this yields
\begin{equation}
\frac{T+T_e^*}{2(T_e^*)^{3/2}} \sqrt{\delta(1-\delta)} \left[ Q^{-1}(\eeff)(1-\eeff) + \frac{1}{\sqrt{2\pi}} e^{-\frac{\left(Q^{-1}(\eeff)\right)^2}{2}} \right] = (1-\delta)(1-\eeff)
\label{eqn:T_e_maximization_expr}
\end{equation}
Numerically solving \eqref{eqn:T_e_maximization_expr} would yield us $T_e^*$ in terms of $T$.

Assuming that $T \gg T_e^*$, we see that for fixed $\eeff$, 
\begin{equation}
    \frac12 T \cdot \sqrt{\frac{\delta (1 - \delta)}{(T_e^*)^3}} \left(Q^{-1}(\eeff)(1 - \eeff) + \frac{1}{\sqrt{2 \pi}} e^{-\frac{(Q^{-1}(\eeff))^2}{2}} \right) - (1 - \delta) \cdot (1 - \eeff) = 0
\end{equation}

\section{Proof of Theorem~\ref{thm:optimum_NNN}}
\label{appendix:appendix_4}

We take \eqref{eqn:NNN_expression} and again neglect the $O\left(\frac{T_t}{T_e^{\frac{3}{2}}} \right)$ terms. Now, we maximize this expression with respect to $\bb$. Differentiating with respect to $\bb$ yields
\begin{IEEEeqnarray}{rCl}
    - (1 - \eeff) + (1 - \delta - \bb) \cdot e^{-\frac{\bb^2 T_e}{2 \delta (1 - \delta)}} \sqrt{\frac{T_e}{2 \pi \delta (1 - \delta)}} + \sqrt{\frac{\delta (1 - \delta)}{2 \pi T_e}} e^{-\frac{\bb^2 T_e}{2 \delta (1 - \delta)}} \left(\frac{T_e \bb}{ \delta (1 - \delta)} \right) &=& 0 \\
    - (1 - \eeff) + (1 - \delta) \cdot e^{-\frac{\bb^2 T_e}{2 \delta (1 - \delta)}} \sqrt{\frac{T_e}{2 \pi \delta (1 - \delta)}} &=& 0 \\
    (1 - \delta) \cdot e^{-\frac{(Q^{-1} (\eeff))^2}{2}} \sqrt{\frac{T_e}{2 \pi \delta (1 - \delta)}}  &=& (1 - \eeff)  
\end{IEEEeqnarray}

This is a function purely of $\eeff$, and thus can be maximized numerically to get a solution.

\section{Proof of Theorem~\ref{thm:geometric_windowing}}
\label{appendix:appendix_5}

First, we rearrange the expression for $\NNN$ from Theorem~\ref{thm:exp_exploration}.

\begin{IEEEeqnarray}{rCl}
    \NNN(\bb, \delta, T, T_e) &=& T_t \cdot \left((1 - \delta - \sqrt{\frac{\delta (1 - \delta)}{T_e}}Q^{-1}({\eeff})) \cdot (1 - \eeff) -\sqrt{\frac{\delta (1 - \delta)}{2 \pi T_e}} e^{-\frac{(Q^{-1} (\eeff))^2}{2}}
    \right) \\
    &=& T_t \cdot \left((1 - \delta) (1 - \eeff) - \sqrt{\frac{\delta (1 - \delta)}{T_e}} \left(Q^{-1}({\eeff}) (1 - \eeff) + \frac{e^{-\frac{(Q^{-1} (\eeff))^2}{2}}}{\sqrt{2 \pi}} \right)
    \right)    
\end{IEEEeqnarray}

For the geometric windowing strategy with blocklengths $T_i = 2^{i - 1}$. For the ith block, $T_t = 2^i$ and $T_e = 2^i - 1$. $T = 2^{M} - 1$. First, we prove the upper bound:
\begin{IEEEeqnarray}{rCl}
    \NNN(\delta, M, \vec T) &=& \sum_{i = 1}^M 2^{i - 1} \cdot \left((1 - \delta) (1 - \eeff) - \sqrt{\frac{\delta (1 - \delta)}{2^{i - 1} - 1}} \left(Q^{-1}({\eeff}) (1 - \eeff) + \frac{e^{-\frac{(Q^{-1} (\eeff))^2}{2}}}{\sqrt{2 \pi}} \right)
    \right) \\
    &\overset{(a)}{\le}& \sum_{i = 1}^M 2^{i - 1} \cdot \left((1 - \delta) (1 - \eeff) - \sqrt{\frac{\delta (1 - \delta)}{2^{i - 1}}} \left(Q^{-1}({\eeff}) (1 - \eeff) + \frac{e^{-\frac{(Q^{-1} (\eeff))^2}{2}}}{\sqrt{2 \pi}} \right)
    \right) \\
    &\overset{(b)}{=}& T (1 - \delta) (1 - \eeff) - \frac{\sqrt{(T + 1) \delta (1 - \delta)}}{\sqrt{2} - 1} \left(Q^{-1}({\eeff}) (1 - \eeff) + \frac{e^{-\frac{(Q^{-1} (\eeff))^2}{2}}}{\sqrt{2 \pi}} \right),
\end{IEEEeqnarray}
where $(a)$ follows since $2^{i - 1} - 1 \le 2^{i - 1}$ and $(b)$ follows by finding the sum of the geometric progression.

Then, we prove the lower bound:
\begin{IEEEeqnarray}{rCl}
    \NNN(\delta, M, \vec T) &=& \sum_{i = 1}^M 2^{i - 1} \cdot \left((1 - \delta) (1 - \eeff) - \sqrt{\frac{\delta (1 - \delta)}{2^{i - 1} - 1}} \left(Q^{-1}({\eeff}) (1 - \eeff) + \frac{e^{-\frac{(Q^{-1} (\eeff))^2}{2}}}{\sqrt{2 \pi}} \right)
    \right) \\
    &\overset{(a)}{\ge}& \sum_{i = 1}^M 2^{i - 1} \cdot \left((1 - \delta) (1 - \eeff) - \sqrt{\frac{\delta (1 - \delta)}{2^{i - 2}}} \left(Q^{-1}({\eeff}) (1 - \eeff) + \frac{e^{-\frac{(Q^{-1} (\eeff))^2}{2}}}{\sqrt{2 \pi}} \right)
    \right) \\
    &\overset{(b)}{=}& T (1 - \delta) (1 - \eeff) - \frac{\sqrt{2(T + 1) \delta (1 - \delta)}}{\sqrt{2} - 1} \left(Q^{-1}({\eeff}) (1 - \eeff) + \frac{e^{-\frac{(Q^{-1} (\eeff))^2}{2}}}{\sqrt{2 \pi}} \right),
\end{IEEEeqnarray}
where $(a)$ follows since $2^{i - 1} - 1 \ge 2^{i - 2}$ and $(b)$ follows by finding the sum of the geometric progression.

\end{document}